\documentclass[twocolumn,showpacs,aps,prl]{revtex4}  
\usepackage{epsfig,amssymb}

\begin{document}

\title{Observation of Faraday Waves in a Bose-Einstein Condensate}

\author{P. \surname{Engels}}
\email{engels@wsu.edu}
\author{C. \surname{Atherton}}
\affiliation{Washington State University, Department of Physics
and Astronomy, Pullman, WA, 99164}
\author{M. A. \surname{Hoefer}}
\affiliation{National Institute of Standards and Technology,
Boulder, Colorado 80305}
\thanks{Contribution of the U.S. Government.  Not subject to copyright.}

\date{\today}

\begin{abstract}
Faraday waves in a cigar-shaped Bose-Einstein condensate are
created. It is shown that periodically modulating the
\textit{transverse} confinement, and thus the nonlinear
interactions in the BEC, excites small amplitude
\textit{longitudinal} oscillations through a parametric resonance.
It is also demonstrated that even without the presence of a
continuous drive, an initial transverse breathing mode excitation
of the condensate leads to spontaneous pattern formation in the
longitudinal direction.  Finally, the effects of strongly driving
the transverse breathing mode with large amplitude are
investigated. In this case, impact-oscillator behavior and
intriguing nonlinear dynamics, including the gradual emergence of
multiple longitudinal modes, are observed.
\end{abstract}

 \maketitle

In 1831, Faraday studied the behavior of liquids that are
contained in a vessel subjected to oscillatory vertical motion
\cite{Faraday1831}. He found that fluids including alcohol, white
of egg, ink and milk produce regular striations on their surface.
These striations oscillate at half the driving frequency and are
termed Faraday waves. They are considered to be an important
discovery. Since then, the more general topic of pattern formation
in driven systems has been met with great interest, and patterns
have been observed in hydrodynamic systems, nonlinear optics,
oscillatory chemical reactions, and biological media
\cite{Cross1993}.
\par
In this paper we study pattern formation by modulating the
nonlinearity in a Bose-Einstein condensate (BEC). Nonlinear
dynamics arise from the interatomic interactions in this ultracold
gas. In the past the observation of interesting phenomena has
motivated researchers to propose and implement various techniques
to manipulate the nonlinearity. Such control has been accomplished
for example by exploiting Feshbach resonances \cite{Inouye1998}.
In our experiment we investigate an alternative technique, namely
periodically modulating the nonlinearity by changing the radial
confinement of an elongated, cigar-shaped BEC held in a magnetic
trap. The radial modulation leads to a periodic change of the
density of the cloud in time, which is equivalent to a change of
the nonlinear interactions and the speed of sound. This can, in
turn, lead to the parametric excitation of longitudinal sound-like
waves in the direction of weak confinement. This process is
analogous to Faraday's experiment where the vertical motion of the
vessel produced patterns that were laterally spread out.
\par
It has been shown theoretically that for a BEC, a Faraday type
modulation scheme in the case of small driving frequencies leads
one to the same type of analysis as would the direct modulation of
the interatomic interaction, e.g., by a Feshbach resonance
\cite{Staliunas2002,Staliunas2004}. In both cases, the dynamics
are governed by a Mathieu equation that is typical for
parametrically driven systems. Floquet analysis reveals that a
series of resonances exist, consisting of a main resonance at half
the driving frequency, and higher resonance tongues at integer
multiples of half the driving frequency \cite{Staliunas2002}.

In our experiment we exploit this transverse modulation scheme for
three different applications. First, we apply a relatively weak
continuous modulation, demonstrate the emergence of longitudinal
Faraday waves, and study their behavior as a function of the
excitation frequency. Second, we investigate longitudinal patterns
that emerge as a consequence of an initial transverse breathing
mode excitation without the presence of a continuous drive. This
has important consequences in the context of damped BEC
oscillations and has been studied theoretically in
\cite{Kagan2001}. Since the first experiments with BECs, the study
of collective excitations has been a central theme
\cite{Jin1996,Mewes1996}. The transverse breathing mode, which we
exploit in our experiments, plays a prominent role: Chevy et al.
\cite{Chevy2002} showed that this mode exhibits unusual
properties, namely an extremely high quality factor and a
frequency nearly independent of temperature. Finally, in a third
set of measurements we study the situation of a relatively strong
modulation, resonantly driving the transverse breathing mode. We
show that the condensate responds as an impact oscillator, which
leads to intriguing multimode dynamics.

 The experiments were carried out in a newly
constructed BEC machine that produces cigar-shaped condensates of
$^{87}$Rb atoms in the $|F=1,m_{F}=-1\rangle$ state. The typical
atom number in the BEC is $5 \cdot10^{5}$, and the atoms are
evaporatively cooled until no thermal cloud surrounding the
condensate is visible any more. The atoms are held in a
cylindrically symmetric Ioffe-Pritchard type magnetic trap with
the harmonic trapping frequencies of
$\{\omega_{xy}/(2\pi),\omega_{z}/(2\pi)\}=\{160.5,7\}$Hz. The
weakly confined z-direction is oriented horizontally. For the
experiments described below, the following collective mode
frequencies are of particular importance: first, there exists a
high-frequency transverse breathing mode. For our trap geometry,
this mode has a frequency of $\omega_{\bot}/(2\pi) = 321$~Hz
\cite{Stringari1996,Garcia1996}, very close to the limit of
vanishing axial confinement $2\cdot\omega_{xy}/(2\pi)$. The second
set of modes in which we are interested here consists of axial
modes which, for large quantum numbers, correspond to sound waves
in the z-direction. The frequencies of this discrete set of modes
can be approximately calculated as given in
\cite{Fliesser1997,Stringari1998}.

\begin{figure} \leavevmode \epsfxsize=3.375in
\epsffile{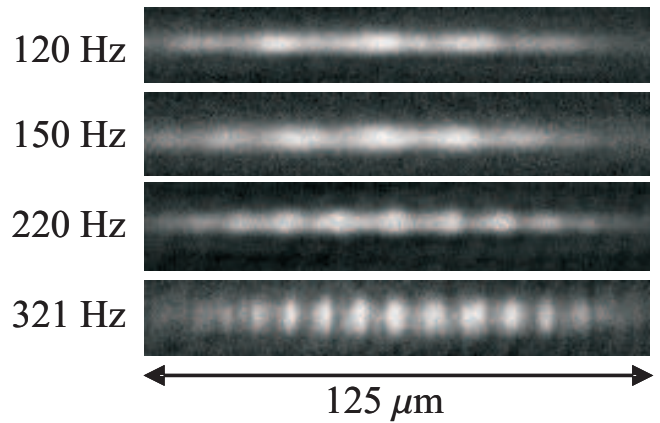} \caption{\label{faradaypatterns} In-trap
absorption images of Faraday waves in a BEC. Frequency labels for
each image represent the driving frequency at which the transverse
trap confinement is modulated.}
\end{figure}

In order to investigate the parametric driving process mentioned
above, we first performed a set of "spectroscopy" experiments in
which we continuously modulated the transverse trapping
confinement at a fixed modulation frequency and observed the
subsequent emergence of longitudinal Faraday waves. For each
excitation frequency the modulation amplitude was adjusted such
that the longitudinal patterns emerged typically at some point
after 10 to 30 oscillations. On the breathing mode resonance, a
trap modulation of 3.6\% was used, while at many other frequencies
trap modulations of up to 42.5\% were chosen to obtain clearly
visible patterns \cite{trapmodulation}. This range of modulation
depths is similar to the range used in numerical simulations in
\cite{Staliunas2004}. Representative examples for the resulting
patterns are shown in Fig.~\ref{faradaypatterns}. All experimental
images in this manuscript were taken by destructive in-trap
imaging.

\begin{figure} \leavevmode \epsfxsize=3.375in
\epsffile{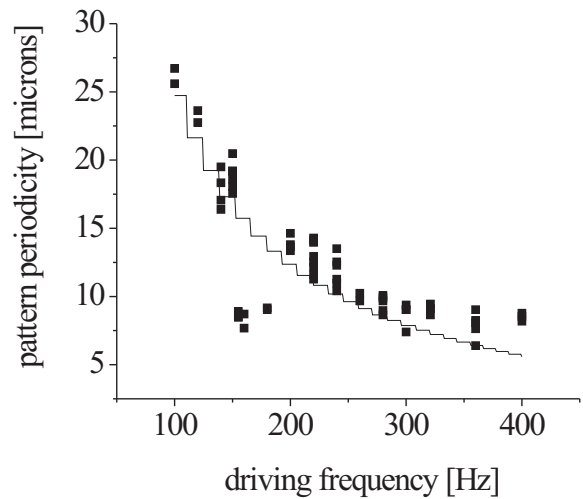} \caption{\label{spectroscopy} Average spacing
of adjacent maxima of the longitudinal patterns plotted versus the
transverse driving frequency. Points are experimental data, while
the line shows the theoretical values calculated for the
longitudinal modes closest to half the driving frequency.}
\end{figure}

The average spacing of adjacent maxima in the resulting pattern is
plotted against the driving frequency in Fig.~\ref{spectroscopy}.
The data lie on a clear curve, with the exception of the points
near a driving frequency of 160.5~Hz, corresponding to the
transverse dipole mode resonance (i.e. transverse slosh
motion)\cite{Stringari1996}. However, inspection of our
experimental images reveals that, at this frequency, we also
excite the transverse breathing mode at 321~Hz. Excitation of the
breathing mode is a very effective way of creating longitudinal
patterns. Therefore the patterns obtained at 160.5~Hz are actually
the same as those produced at 321~Hz. In order to rationalize the
data, we first note that parametric excitation with a certain
driving frequency excites oscillations predominantly at half the
driving frequency, the main resonance also observed in Faraday's
experiments. The dispersion relation of longitudinal collective
modes that become sound-like for high quantum numbers is given in
\cite{Fliesser1997,Stringari1998} and is used to calculate the
expected spacing between density maxima. The resulting spacings
are plotted as the step-like curve in Fig.~\ref{spectroscopy} and
are in excellent agreement with our experimental data,
corroborating the assumption and theory of a parametric driving
process.

\begin{figure} \leavevmode \epsfxsize=3.375in
\epsffile{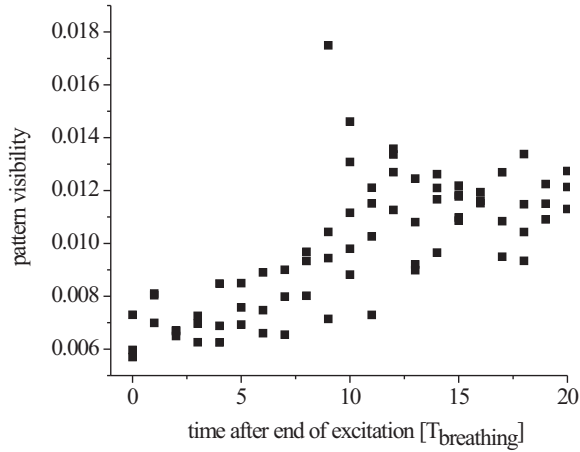} \caption{\label{onset} Onset of Faraday wave
formation after driving the transverse breathing mode for only two
cycles. Pattern visibility is quantified as described in
\cite{Patternquantify}.}
\end{figure}

In a second set of experiments we show that the longitudinal modes
are driven by the transverse breathing motion even without the
presence of a continuous external drive. In particular, this
disproves the influence of any tiny residual axial trap modulation
that has existed during the presence of our continuous drive due
to experimental imperfections \cite{trapmodulation}. For this, we
excited the transverse breathing mode at 321~Hz by driving the
transverse trap confinement for a few cycles, and then let the
condensate evolve without the presence of the drive. For a gentle
excitation, we do not see longitudinal patterns immediately after
the end of the drive, but can observe them emerge at later times.
A weaker excitation delays the emergence of longitudinal modes out
to later times, while in the case of strong excitations the
patterns can emerge within the first three cycles. In order to
follow the evolution, we quantify the presence of longitudinal
patterns as described in \cite{Patternquantify}. Fig.~\ref{onset}
shows the evolution after the end of a moderate excitation. For
this data, we excited the condensate for only two cycles, varying
the transverse trap frequency by 9\% with a modulation frequency
of 321~Hz during the modulation. Immediately after the excitation,
the obtained images showed no longitudinal waves, and our pattern
visibility measure, plotted in the figure, is initially picking up
high frequency noise along the longitudinal axis. A main
contribution to this noise is the imaging noise of our detection
system. Weak pattern formation is observed starting at five
periods after the end of the modulation, and strong longitudinal
patterns then appear after about nine periods. A similar behavior
is known for example from parametric amplifiers in optics: if no
input signal amplitude is present, a signal emerging from noise
(or zero point energy) can form if the amplification is large
enough. This behavior is also found by our numerical simulations
based on the Gross-Pitaevskii equation. In simulations with
similar parameters as in the experiment, no pattern formation is
observed for several breathing mode periods. From the onset of
pattern formation, it takes just three periods for patterns to
grow to their full strength. We find that the onset time of
pattern formation in the simulations is earlier when more noise is
added to the initial relaxed wave function.

\begin{figure} \leavevmode \epsfxsize=3.375in
\epsffile{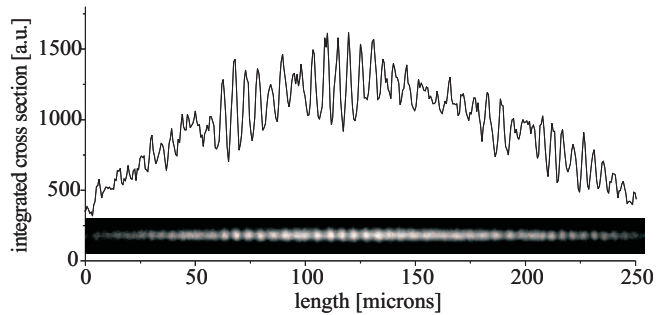} \caption{\label{domains} Pattern domains
observed in a very elongated BEC. The graph shows an integrated
cross section of the BEC displayed in the inset.}
\end{figure}

This experiment is closely related to the theoretical situation
described in \cite{Kagan2001}, where a single and sudden jump in
the transverse trap frequency was used to excite the breathing
mode, instead of a sinusoidal trap frequency modulation. We have
used a trap jumping excitation in the case of very elongated BECs
to demonstrate a second observation about the onset of
longitudinal patterns, namely the fact that these patterns can
start to emerge in spatially localized domains, rather than
uniformly across the whole cloud. To show this effect, we produced
condensates in a very elongated cigar-shaped trap with trapping
frequencies of
$\{\omega_{xy}/(2\pi),\omega_{z}/(2\pi)\}=\{286.1,2.8\}$Hz. We
temporarily jumped to a different trap of
$\{\omega_{xy}/(2\pi),\omega_{z}/(2\pi)\}=\{88.4,5.1\}$Hz for the
duration of 1.3~ms, then jumped back to the first trap and let the
cloud evolve. After about 10~ms, we observed BECs in which a
perfectly periodic density modulation was stretching over almost
the entire cloud; but it is also not uncommon to observe patterns
in several separate domains, as shown in Fig.~\ref{domains}.
Considering that the speed of sound in the elongated cloud is only
about 2~mm/s \cite{Zaremba1998,Stringari1998}, it is plausible
that over the time scale of this experiment different regions of
the BEC can get independently excited and evolve into longitudinal
patterns independently from each other.

\begin{figure} \leavevmode \epsfxsize=3.375in
\epsffile{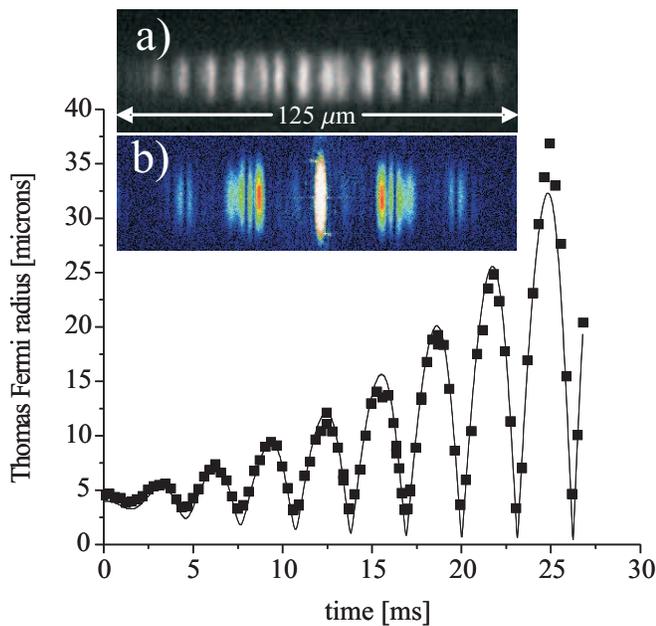} \caption{\label{impactosci} Impact oscillator
behavior. Plot of BEC radius in the tight direction vs. time when
continuously driving the transverse breathing mode. Points show
experimental data, while the curve corresponds to numerical
simulation. Inset: (a) Image of BEC after 5.2 driving periods,
i.e. 16.2~ms. (b) Fourier transform of the image in (a).}
\end{figure}

In the experiments described so far, our parametric excitation has
led to longitudinal modes oscillating at half the frequency of the
parametric driving frequency. In the case of strong parametric
amplification, it is theoretically expected that modes at other
frequencies (higher resonance tongues), in particular modes at the
driving frequency, can be excited, too \cite{Staliunas2002}. This
motivated our third set of experiments in which we started again
with a condensate in a trap with trap frequencies of
$\{\omega_{xy}/(2\pi),\omega_{z}/(2\pi)\}=\{160.5,7\}$Hz.  We then
continuously modulated our transverse trap frequency by about 19\%
with a modulation frequency of 321~Hz. The resulting breathing
motion is seen in Fig.~\ref{impactosci} where we plot the
Thomas-Fermi radius of the cloud in the transverse direction
versus time. The graph clearly shows that the cloud, upon strong
excitation, gradually starts behaving as an impact oscillator,
i.e., as an oscillator bouncing off a stiff wall during each
period. A classical impact oscillator, realized for example by a
ball bouncing off a stiff surface, is a paradigm for nonlinear
dynamics, nonlinear resonances and chaotic behavior. In the
present case, the role of the stiff wall was played by the strong
mean-field repulsion during the slim phase of the oscillations
when the BEC is strongly compressed in the transverse direction.
Theoretically, the instability of the breathing mode upon strong
driving and the impact oscillator behavior have been analyzed in
\cite{Ripoll1999}. The behavior of the cloud radius in the
transverse direction was also reproduced in our numerical
simulation of the azimuthally symmetric 3D Gross-Pitaevskii
equation, displayed as the solid line in Fig.~\ref{impactosci}.
However, the numerics, when starting with a thoroughly relaxed
wave function in the initial trap, show a sign of longitudinal
pattern formation only after 18~ms, while in the experiment,
longitudinal patterns clearly formed already during the third
period (9~ms). This, again, hints at the importance of initial
noise in the condensate that seeds the parametric amplification.
In the experiment, the patterns start out similar to those
displayed in Fig.~\ref{faradaypatterns} for the case of a weak
drive at 321~Hz. But, upon the action of the strong drive, they
quickly evolve into more complicated patterns, involving the
excitation of several other modes. The inset of
Fig.~\ref{impactosci} shows an image taken after 5.2 driving
periods (a), together with the Fourier transform (b). The Fourier
spectrum reveals that several modes corresponding to the first
resonance tongue of longitudinal modes with nearly half the
driving frequency are excited. In addition, modes at twice the
distance from the central Fourier peak are visible. Those modes
belong to the second resonance tongue of the main resonance.

\par
In conclusion, we have experimentally observed the effects of
parametric resonances in a BEC. The observed resonances lead to
Faraday waves along the long BEC axis. These results advance the
understanding of collective mode behavior in a condensate, which
is one of the key tools to study BEC dynamics. In addition, we
have shown that strongly driving the transverse breathing mode
leads to an instability whereupon the mode amplitude increases
exponentially, accompanied by the strong excitation of multiple
sound-like modes.

\begin{acknowledgments}
\end{acknowledgments}


\begin{thebibliography}{Hi!}
\bibitem{trapmodulation}
Sinusoidally modulating our bias field strength led to a
modulation of the transverse confinement $\omega_{xy}/(2\pi)$
between 155~Hz and 166.7~Hz in the case of a 3.6\% modulation
depth, and between 122~Hz and 303~Hz for our strongest modulation
depth of 42.5\%. Even with our strongest transverse modulation,
the longitudinal confinement was modulated by less than 0.6\%.
\bibitem{Patternquantify}
We integrate the observed density along the tight direction, and
determine the standard deviation of the resulting cross section
from a Thomas-Fermi shape in the central region. We normalize the
result by dividing by the number of atoms in the same region. The
Thomas-Fermi shape is obtained by eliminating all but the low
frequency Fourier components that correspond to the overall shape
of the BEC. This procedure provides a good measure for the
excitation of the longitudinal modes, which we have checked also
by visual inspection of the data.
\end{thebibliography}
\end{document}